\documentclass{emulateapj}
\usepackage{multirow}
\usepackage{subfigure}
\def\msun{$M_{\odot}$}

\def\xmm{{\it XMM-Newton}}

\shortauthors{Lin et al.}
\begin{document}

\title{Discovery of a Highly Variable Dipping Ultraluminous X-ray source in M94}

\author{Dacheng Lin\altaffilmark{1,2,3}, Jimmy A. Irwin\altaffilmark{1}, Natalie A. Webb\altaffilmark{2,3},  Didier Barret\altaffilmark{2,3}, and Ronald A. Remillard\altaffilmark{4}}
\altaffiltext{1}{Department of Physics and Astronomy, University of Alabama, Box 870324, Tuscaloosa, AL 35487, USA, email: dlin@ua.edu}
\altaffiltext{2}{CNRS, IRAP, 9 avenue du Colonel Roche, BP 44346, F-31028 Toulouse Cedex 4, France}
\altaffiltext{3}{Universit\'{e} de Toulouse, UPS-OMP, IRAP, Toulouse, France}
\altaffiltext{4}{MIT Kavli Institute for Astrophysics and Space Research, MIT, 70 Vassar Street, Cambridge, MA 02139-4307, USA}

\begin{abstract}
We report the discovery of a new ultraluminous X-ray source (ULX) \object{2XMM J125048.6+410743} within the spiral galaxy M94. The source has been observed by {\it ROSAT}, {\it Chandra}, and {\it XMM-Newton} on several occasions, exhibiting as a highly variable persistent source or a recurrent transient with a flux variation factor of $\gtrsim$100, a high duty cycle (at least $\sim$70\%), and a peak luminosity of $L_{\rm X}$$\sim$2$\times$10$^{39}$ erg~s$^{-1}$ (0.2--10 keV, absorbed). In the brightest observation, the source is similar to typical low-luminosity ULXs, with the spectrum showing a high-energy cutoff but harder than that from a standard accretion disk. There are also sporadical short dips, accompanied by spectral softening. In a fainter observation with $L_{\rm X}$$\sim$3.6$\times$10$^{38}$ erg~s$^{-1}$, the source appears softer and is probably in the thermal state seen in Galactic black-hole X-ray binaries (BHBs). In an even fainter observation ($L_{\rm X}$$\sim$9$\times$10$^{37}$ erg~s$^{-1}$), the spectrum is harder again, and the source might be in the steep-powerlaw state or the hard state of BHBs. In this observation, the light curve might exhibit $\sim$7 hr (quasi-)periodic large modulations over two cycles. The source also has a possible point-like optical counterpart from {\it HST} images. In terms of the colors and the luminosity, the counterpart is probably a G8 supergiant or a compact red globular cluster containing $\sim2\times10^5$ K dwarfs, with some possible weak UV excess that might be ascribed to accretion activity. Thus our source is a candidate stellar-mass BHB with a supergiant companion or with a dwarf companion residing in a globular cluster. Our study supports that some low-luminosity ULXs are supercritically accreting stellar-mass BHBs.

\end{abstract}

\keywords{accretion, accretion disks --- black hole physics --- X-rays: binaries --- X-rays:individual:\object{2XMM J125048.6+410743}}

\section{INTRODUCTION}
\label{sec:intro}
Many bright X-ray sources have been discovered in nearby
galaxies. Those with luminosities $L_{\rm X}$ exceeding 10$^{39}$
erg~s$^{-1}$, the Eddington limit for a black-hole X-ray binary (BHBs)
with a black-hole (BH) mass of $M_{\rm BH}\sim10$ \msun, are often
referred to as ultraluminous X-ray sources \citep[ULXs, for a recent
  review see][]{feso2011}. While a very small number of them appear as
hyperluminous X-ray sources (HLXs, especially \object{ESO 243-49
  HLX-1}) with $L_{\rm X}>10^{41}$ erg~s$^{-1}$ and are strong
candidates for intermediate-mass BHs (IMBHs) with $M_{\rm
  BH}\sim10^2$--$10^5$ \msun\ \citep{faweba2009, jotofa2010,
  sefali2011, surowa2012,goplka2012}, most ULXs have $L_{\rm
  X}<10^{40}$ erg~s$^{-1}$. The nature of these 'low-luminosity' ULXs
is still under strong debate. They could be IMBHs
\citep[e.g.,][]{comu1999}, super-Eddington low-mass BHs
\citep[e.g.,][]{kosc1998}, beamed emission from geometrically thick
disks at high accretion rates \citep{kidawa2001} or from relativistic
jets \citep{okdoma1998}, or combination of these effects
\citep{polifa2007,ki2009}.

Due to their large distances, there are no confident dynamical mass
measurements for ULXs yet \citep{feso2011}, unlike Galactic BHBs,
whose dynamical masses of the BHs are constrained to below 20
\msun\ \citep{remc2006}. The spectral and timing properties of
Galactic BHBs have been well studied, and they are known to exhibit
three characteristic X-ray spectral states: the thermal state has a
dominant thermal disk and weak fast variability, typically at
luminosities above a few percent of the Eddington limit; the hard
state has a dominant hard powerlaw (PL, with photon index $\Gamma_{\rm
  PL}$ typically within 1.4--2.1) and strong fast variability,
generally seen at low luminosities; and the steep-PL state has a
powerful PL with $\Gamma_{\rm PL}\sim 2.5$ and commonly-occuring
quasi-periodic oscillations, often seen at very high luminosities
\citep{remc2006,mcre2006,dogiku2007}. These states also often show
correlated radio emission \citep{febega2004}. Thus studies of ULXs
often compare their properties with those of Galactic BHBs to infer
their possible nature.

Two transient ULXs were discovered in M31 recently
\citep{misuro2012,mimima2013}. When they reached luminosities above
10$^{39}$ erg~s$^{-1}$, their X-ray spectra appeared harder than those
from a standard geometrically thin optically thick disk. This is often
seen in low-luminosity ULXs, and one possible explanation is Compton
up-scattering of disc photons in a wind or photosphere
\citep{misuro2012,mimima2013}. The above similarity led
\citet{misuro2012,mimima2013} to conclude that these two transient
ULXs in M31 are members of low-luminosity ULXs, though the latter are
mostly persistent with small long-term variability
\citep{feso2011}. However, these two ULXs are also similar to BHBs
with stellar-mass BHs in other aspects, such as the outburst profile
and the coupled X-ray and radio behavior. Thus
\citet{misuro2012,mimima2013} concluded that many low-luminosity ULXs
are stellar-mass BHBs.

In \citet{liweba2012}, we classified 4330 sources from the 2XMMi-DR3
catalog \citep{wascfy2009}. In this project, many sources showing
interesting behavior but poorly studied in the literature were also
discoverd. We are devoting a series of papers
\citep[e.g.,][]{licagr2011,liweba2013b,liweba2013} to present the
properties of these sources in detail. Here we continue our study and
concentrate on a highly variable ULX \object{2XMM J125048.6+410743} in
the spiral galaxy M94 (NGC 4736), which shows some evidence of dips,
periodic oscillations, and a state transition, and has a possible
optical counterpart from the \textit{Hubble Space Telescope}
(\textit{HST}) images. The source, whose 1$\sigma$ positional error is
0\farcs35 per coordinate from the 2XMMi-DR3 catalog, is only 0\farcs26
away from \object{CXO J125048.6+410742} (RA=12h50m48.605s,
Dec=+41d07$\arcmin$42.35$\arcsec$, and the 95\% error is 0\farcs29) in
the {\it Chandra} Source Catalog \citep[CSC, release
  1.1,][]{evprgl2010}. Thus we conclude that they are the same
source. It should also be the off-nuclear source detected in a {\it
  ROSAT}/PSPCB observation in \citet{cufebr1997}, because they
coincide within 5$\arcsec$ in position and there are no other bright
sources within 20$\arcsec$ from {\it Chandra} observations. We
designate the source as ULX2 hereafter, considering that there is a
different ULX \citep[U30 in][]{beweco2008}, which is hardly variable
and has a flux comparable to the peak value of our source. In
Section~\ref{sec:reduction}, we describe the analysis of X-ray and
optical observations. In Section~\ref{sec:res}, we present the
results. In Section~\ref{sec:dis}, we discuss the nature of the source
and the cause of its short-term X-ray variability. The conclusions of
our studies are given in Section~\ref{sec:conclusion}. Throughout the
paper, we assume a source distance of 5.2 Mpc \citep{todrbl2001}.

\section{DATA ANALYSIS}
\label{sec:reduction}

\subsection{X-ray Observations}
\label{sec:redxrayobs}

\tabletypesize{\scriptsize}
\setlength{\tabcolsep}{0.02in}
\begin{deluxetable}{rccccccccc}
\tablecaption{The X-ray Observation Log\label{tbl:obslog}}
\tablewidth{0pt}
\tablehead{\colhead{Obs. ID} &\colhead{Date} & \colhead{Detector} &\colhead{OAA} &\colhead{$T$ (ks)} &\colhead{$r_{\rm src}$} &\colhead{S/N}  \\
(1) & (2) &(3) & (4) & (5) & (6) & (7)
}
\startdata
\multicolumn{4}{l}{\xmm:}\\
\hline
0094360601(X1) &2002-05-23 & pn/M1/M2 & 2.3$\arcmin$ & 15/20/20 & 15$\arcsec$ & 5 \\
0094360701(X2) &2002-06-26 & pn/M1/M2 & 2.5$\arcmin$ & 4/12/12 & 15$\arcsec$ & 5 \\
0404980101(X3) &2006-11-28 & pn/M1/M2 & 1.6$\arcmin$ & 34/46/46 & 15$\arcsec$ & 112 \\
\hline
\multicolumn{4}{l}{{\it Chandra}:}\\
\hline
808(C1)  & 2000-05-13 & ACIS-S3 & 0.8$\arcmin$  & 47 & 1.5$\arcsec$ & 17 \\
9553(C2)  & 2008-02-16 & ACIS-I2 & 0.5$\arcmin$  & 24 & 1.5$\arcsec$ & 18 \\
\hline
\multicolumn{4}{l}{{\it ROSAT}:}\\
\hline
RP600050N00(R1) & 1991-06-05 & PSPCB & 1.2$\arcmin$ & 91 & 20$\arcsec$ & 32 \\
RH600678N00(R2) & 1994-12-07 & HRI   & 0.9$\arcmin$ & 112 & 7$\arcsec$ & 1 \\
RH600769N00(R3) & 1994-12-25 & HRI   & 0.9$\arcmin$ & 27 & 7$\arcsec$ & 0
\enddata 
\tablecomments{Columns: (1) the observation ID with our designation given in parentheses, (2) the observation start date, (3) the detector, (4) the off-axis angle, (5) the exposures of data used in final analysis, (6) radius of the source extraction region, (7) the signal to noise ratio of the source, combining all detectors.}
\end{deluxetable}

\tabletypesize{\scriptsize}
\setlength{\tabcolsep}{0.02in}
\begin{deluxetable*}{lccccccccccc}
\tablecaption{The {\it HST}/WFPC2 Observation Log\label{tbl:hst}}
\tablewidth{0pt}
\tablehead{\colhead{Exposure ID} &\colhead{Date} & \colhead{Chip \& Filter} &\colhead{$T$} & S$/$N &\multicolumn{3}{c}{VEGA Mag}   \\
\cline{6-8}
    &     &      &  (s)   &      & Counterpart & G8I & K3V \\
(1) & (2) & (3)  & (4) &  (5) & (6)& (7) & (8) 
}
\startdata
u96r060[1$|$2]m & 2005-05-24 & WF3/F336W & 1800 & 6 &  23.18$\pm$0.18 & 24.65 & 23.93 \\
u6ea800[1r$|$2m] & 2001-07-02 &  WF3/F450W & 460 & 24 & 22.88$\pm$0.05 & 23.05 & 22.96\\
u96r060[3$|$4]m & 2005-05-25 & WF3/F555W & 400 & 40 & 22.25$\pm$0.03 & 22.16 & 22.21\\
u671260[1-5]r & 2001-04-21 & WF2/F656N & 1700 & 8 & 20.81$\pm$0.13 & 21.13 & 21.20\\
u6ea800[3$|$4]r & 2001-07-02 & WF3/F814W & 460 &46 & 21.04$\pm$0.02 & 21.07 & 21.05
\enddata 
\tablecomments{Columns: (1) the exposure ID of the data set, (2) the observation date, (3) the chip and filter, (4) the total exposure, (5) the signal to noise ratio, (6) the VEGA mag of the optical counterpart to ULX2, (7--8) magnitudes estimated based on a G8I stellar spectrum and a K3V stellar spectrum, respectively.}
\end{deluxetable*}

M94 has been observed by many X-ray observatories. However, because
ULX2 is faint and is only 1$\arcmin$ away from the nucleus around
which there are strong diffuse emission and many bright point sources,
including the low-ionization-nuclear-emission-region (LINER) nucleus,
we only studied the eight observations (Table~\ref{tbl:obslog}) from
{\it XMM-Newton}, {\it Chandra} and {\it ROSAT}, thanks to their
relatively large effective areas and high angular resolutions. We
designate their observations as X1-X3, C1-C2, and R1-R3, respectively
(refer to Table~\ref{tbl:hst}). In the {\it XMM-Newton} observations,
all the three European Photon Imaging Cameras, i.e., pn, MOS1 (or M1),
and MOS2 (or M2) \citep{jalual2001,stbrde2001,tuabar2001} were
active. We used SAS 12.0.1 and the calibration files of 2013 January
for reprocessing the X-ray event files and follow-up analysis. The
data in strong background flare intervals are excluded following the
SAS thread for the filtering against high backgrounds. The event
selection criteria followed the default values in the pipeline
\citep[see Table~5 in][]{wascfy2009}. That is, for pn spectra, we used
events with PATTERN $\le4$ and FLAG $=0$ (we also extracted pn light
curves, which used events with (FLAG \& 0xfffffef) $=0$), and for MOS
spectra, we used events with PATTERN $\le12$ and (FLAG \& 0xfffffeff)
$=0$. The two {\it Chandra} observations used the imaging array of the
AXAF CCD Imaging Spectrometer \citep[ACIS; ][]{bapiba1998}. ULX2 falls
in the back-illuminated chip S3 and the front-illuminated chip I2 in
observations C1 and C2, respectively. We analyzed the data with the
CIAO (version 4.5) package and the latest calibration (CALDB
4.5.5.1). The {\it ROSAT} \citep{tr1982} observations were reduced
with FTOOLS 6.12.

We extracted the source emission from a circular region with the
radius for each observation given in Table~\ref{tbl:obslog}. The
background in the XMM-Newton and ROSAT observations strongly depends
on the distance to the nucleus due to the presence of bright sources
near the nucleus and the relatively large point spread functions
(PSFs) of these two observatories, and we estimated it using four
circular regions with the same radius and the same distance to the
nucleus as the source region. The {\it Chandra} observations have no
such problem thanks to their supreme angular resolutions, and we used
a single circular region of 10$\arcsec$ radius near the source to
estimate the background. We extracted both the source and background
spectra/light curves from the source and background regions,
respectively. The response files were constructed for the spectral
fitting. We note that we took special care to construct the response
file for the {\it ROSAT}/PSPCB observation R1. The problem is that we
used a small source extraction region (a radius of 20$\arcsec$) in
order to minimize the contamination from bright sources near the
nucleus, leading to significant PSF loss, which is energy
dependent. The response file (we used the default on-axis version in
the HEASARC archive) does not take into account the PSF loss, and
there is no FTOOLS tool to correct for this. We estimate the PSF loss
using the ROSAT observation RP700130N00 of Mkn 501 (it used the same
gain epoch and has a similar (low) column density and photon index
($\sim$2.6) as our source), and correct the count rate in each channel
of the spectrum of our source. Such a correction is probably very
approximate. Thus the spectral fits to this spectrum might be subject
to some systematic error due to this and will be mostly used to infer
the luminosity.

We rebinned the spectra to have at least 20 counts in each bin so as
to adopt the $\chi^2$ statistic for spectral fits. All spectral models
used include the absorption described by the WABS model in XSPEC, with
the lower limit of $N_{\rm H}$ set to be the Galactic line-of-sight
value of 10$^{20}$ cm$^{-2}$ \citep{kabuha2005}.

To help to identify the optical counterpart
(Section~\ref{sec:hstdata}), we used X-ray sources within 2$\arcmin$
from the nucleus to improve the relative astrometry between {\it
  Chandra} and {\it HST}. To have good relative positions for these
X-ray sources, we carried out the source detection using the CIAO task
\verb|wavdetect| \citep{frkaro2002}. We used only the {\it Chandra}
observation C1 so that all sources of interest are in a single CCD,
which is not the case for observation C2. We performed the detection
on the 0.3--8 keV image, which was binned at a subpixel resolution
(1$/$8 sky pixel) to improve the source positions.

\subsection{{\it HST} Observations}
\label{sec:hstdata}
\begin{figure*} 
\centering
\includegraphics[width=0.98\textwidth]{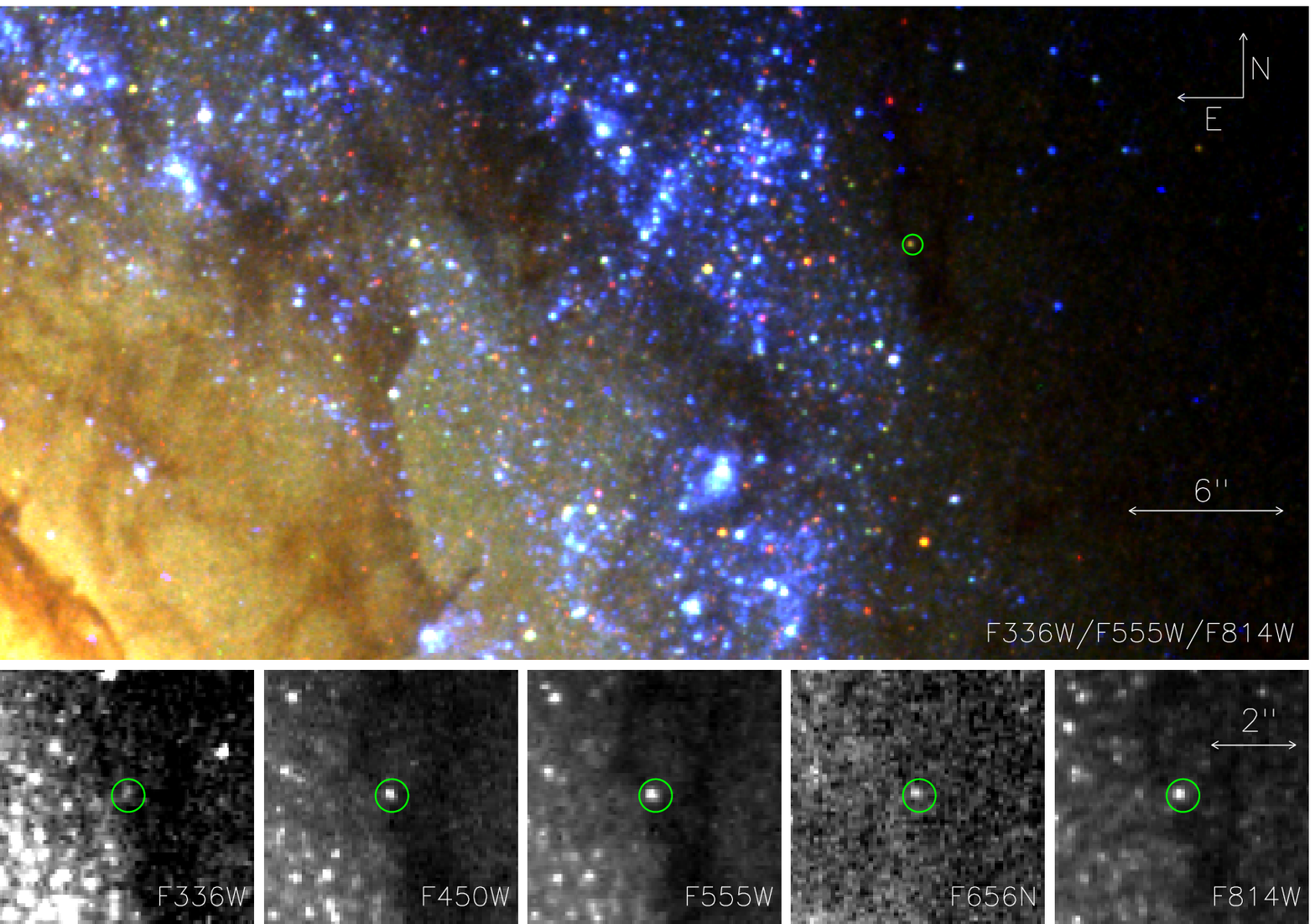}
\caption{The {\it HST}/WFPC2 images around ULX2. The top panel is false-colored using the F814W (red), F555W (green) and F336W (blue) images. The bottom panels, one for each of the five filters, are zoomed in on ULX2. The green circle, with a radius of 0$\farcs$39 (the 95\% systematic error of {\it Chandra} \citep{robu2011}), is centered at the X-ray position, indicating the presence of a possible optical counterpart to ULX2. \label{fig:hstimg}}
\end{figure*}

The sky region around ULX2 was observed by the Wide Field and
Planetary Camera 2 (WFPC2) aboard the {\it HST} several times
(Table~\ref{tbl:hst}). We aligned their drizzled images by matching
the point sources to those in the F555W image, resulting in the rms
deviations $<$0.04$\arcsec$ in all cases. To map the X-ray positions
onto the {\it HST} images, we assumed that \object{CXO
  J125053.0+410713} in the CSC and the UV source at the center of the
{\it HST} images are the nuclear source and coincident with each other
in position \citep[see, e.g.,][]{cose2012}. Because the nuclear UV
source is stronger and less subject to stellar contamination at
shorter wavelength and thus appears more clear in the F336W image than
in the other images, we used the F336W image to determine its position
with the astrolib IDL procedure \verb|cntrd|. We found that it happens
to be only 0\farcs02 away from the position of \object{CXO
  J125053.0+410713} determined by us
(Section~\ref{sec:redxrayobs}). After correcting for this small
relative astrometry, we found an optical point source that is only
0$\farcs$1 away from ULX2 and was detected in all filters
(Figure~\ref{fig:hstimg}). We also found three other X-ray sources
with optical counterpart candidates within 0$\farcs$1, supporting our
astrometry correction. We performed the photometry on the c0f images
with the HSTPHOT 1.1 package \citep{do2000}, which outputs
aperture-corrected VEGA magnitudes.

\section{RESULTS}
\label{sec:res}

\begin{figure} 
\centering
\includegraphics[width=0.48\textwidth]{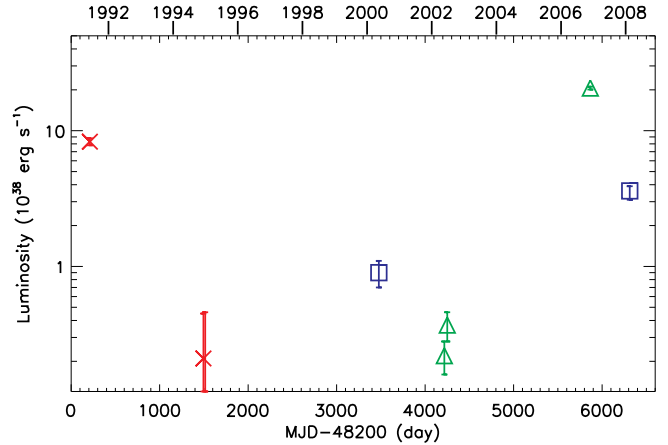}
\caption{The long-term luminosity curve (0.2--10 keV, absorbed), with {\it ROSAT} (red crosses), \xmm\ (green triangles), and {\it Chandra} (blue squares) observations. We note that the third \textit{ROSAT} observation (R3), with $L_\mathrm{x}=0^{+0.46}\times10^{38}$ erg~s$^{-1}$, has only the error bar seen and is very close to the second one (R2, both in 1994 December). \label{fig:srcltlc}}
\end{figure}


\begin{figure*} 
\centering
\subfigure{
\includegraphics[width=0.45\textwidth]{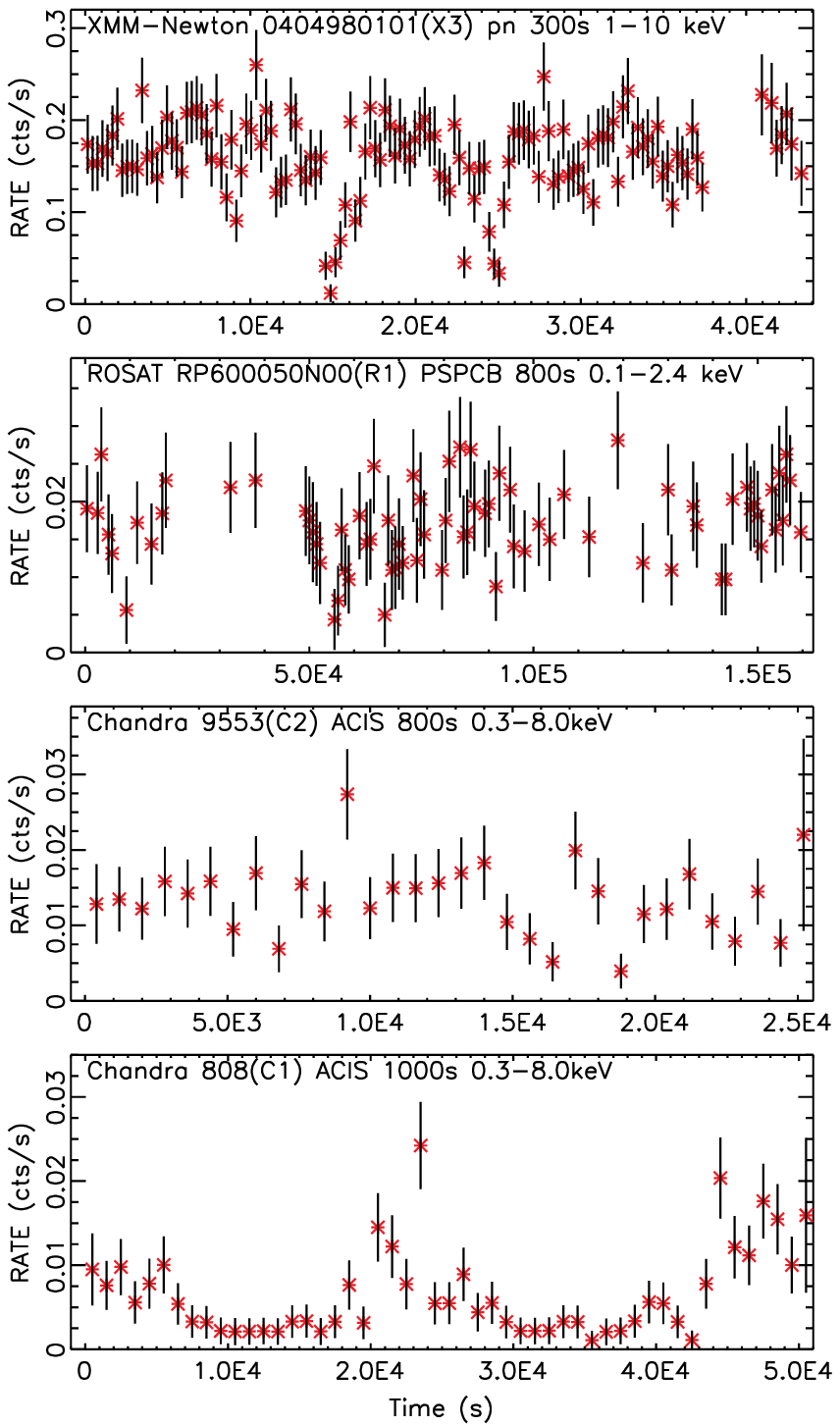}
}
\subfigure{
\includegraphics[width=0.44\textwidth]{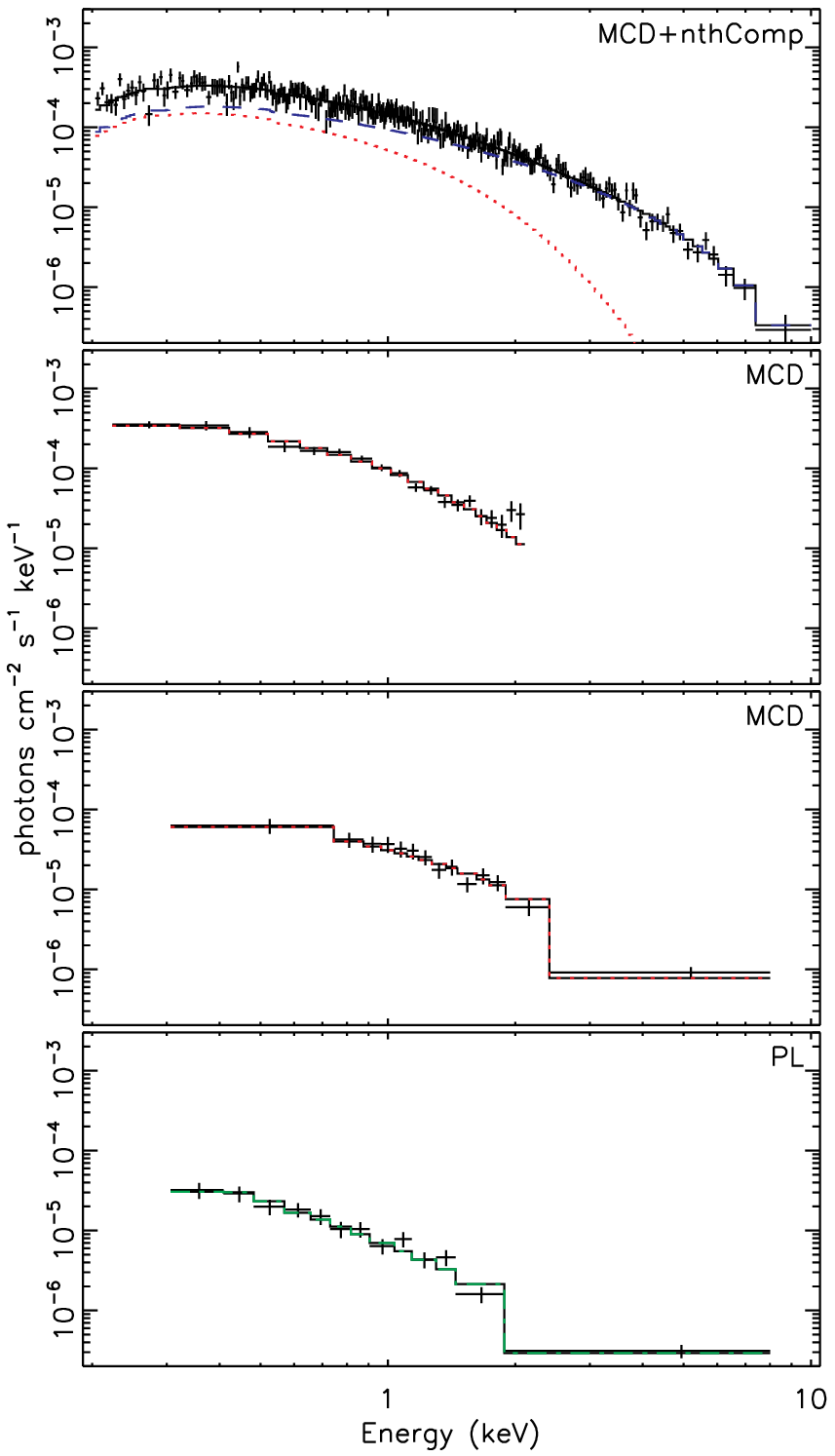}
}
\caption{The light curves (left panels) and unfolded spectra (right
  panels) of the four observations (the luminosity decreases from the
  top to the bottom panels) when ULX2 was significantly detected. The
  notation in the left panels includes the observatory, the
  observation ID, the instrument, the light curve bin size, and the
  energy band used, and that in the right panels is the spectral model
  used. The total model is shown as a black solid line, the MCD
  component as a red dotted line, the nthComp component as a blue
  dashed line, and the PL component as a green dot-dashed
  line.  \label{fig:indlcsp}}
\end{figure*}

\subsection{Long-term X-ray Variability}
Table~\ref{tbl:obslog} lists the signal to noise ratio (S/N) of ULX2
in each observation. The source was significantly detected in
observations X3, C1, C2, and R1 (S/N$\ge$17), but it was only
marginally detected in observations X1 and X2 (S/N$=$5) and was not
detected in observations R2 and R3 (S/N$\le$1).
Figure~\ref{fig:srcltlc} plots the long-term 0.2--10 keV absorbed
luminosity curve of ULX2. The spectral models that we assumed to
derive the luminosities are those shown in Figure~\ref{fig:indlcsp}
for the bright observations (X3, C1, C2 and R1) and a PL for the other
faint observations, which will be described in
Section~\ref{sec:spmod}. The luminosity reached $L_{\rm X}\sim
2\times10^{39}$ erg~s$^{-1}$ in observation X3, supporting ULX2 as a
low-luminosity ULX. This luminosity is a factor of $\sim$100 of that
of observation X1 (we did not use observations R2 and R3 for
comparison because their luminosities have larger
uncertainties). Considering that observations R2 and R3 are close in
time and that X1 and X2 are close in time too, ULX2 was observed in
essentially six epochs and detected in four (or in five if the
detections in X1 and X2 are really from ULX2) of them over nearly two
decades. Thus the source has a large duty cycle ($\sim$70\% or more)
and is probably a highly variable persistent source or a recurrent
transient with probably at least three outbursts in the last two
decades (see Figure~\ref{fig:srcltlc}). Considering that observations
X3 and C2 are about one year apart, the outburst should last more than
one year if they are in the same outburst.

\subsection{Short-term X-ray Variability}
The left panels of Figure~\ref{fig:indlcsp} show the light curves of
the four observations when ULX2 was significantly detected, in order
of decreasing luminosity from the top to the bottom panels. The
brightest observation X3 seems to experience dipping behavior
sporadically (15 ks and 25 ks into the observation), with duration a
few hundred seconds. To understand the spectral properties in the
dipping periods, we extracted the spectrum in the dipping period when
the pn 1-10 keV count rate is $<$0.06 counts~s$^{-1}$ (refer to
Figure~\ref{fig:indlcsp}). The spectrum turns out to be soft. When we
fitted it with a multi-color disk (MCD, diskbb in XSPEC), we obtained
the inner disk temperature $kT_{\rm MCD}=0.36\pm0.06$ keV (the column
density is $N_{\rm H}=1^{+5}\times10^{20}$ cm$^{-2}$, and the reduced
$\chi^2$ is relatively high (2.0) for 6 degrees of freedom
(d.o.f.)). In comparison, the total spectrum gives $kT_{\rm
  MCD}=0.86\pm0.02$ keV (Section~\ref{sec:spmod}). The hardness ratio,
defined as the pn count rate in 1--10 keV divided by that in 0.2-1
keV, is 0.31$\pm$0.06 in the above dipping period, significantly lower
than that (1.05$\pm$0.02) from the whole observation. We also
calculated the ratio of the pn count rate in the dipping period to
that from the whole observation and obtained $0.78\pm0.08$ using
0.2--1 keV events and $0.23\pm0.04$ using 1--2 keV events. Thus the
dips are accompanied by spectral softening. Because of this, we
created the light curve using a high-energy band (1--10 keV) for this
bright observation to show the dips more clearly. Dips might also be
present in the two fainter observations R1 and C2, but not clearly
seen due to low counting statistics.

In contrast, the light curve of the fainter observation C1 might show
three flares that could possibly indicate $\sim$7-hr (quasi-)periodic
modulations over two cycles. Using a 60 ks window to calculate the
count rates at different times, we found that the count rates in the
three flares are about 3.3, 5.0 and 6.8 times of those in the minima
(the count rates in the two minima are similar), corresponding to
about $4\sigma$, $6\sigma$, and $7\sigma$, respectively. We also
extracted the spectra in the bright and faint intervals, depending on
whether the count rate is larger or smaller than 0.005
counts~s$^{-1}$. When we fitted them with a PL, we obtained their
photon indices to be $\Gamma_{\rm PL}$=2.5$\pm$0.2 and 2.6$\pm$0.4,
respectively (we fixed $N_{\rm H}$ at $6\times10^{20}$ cm$^{-2}$
obtained from the fit to the total spectrum (Section~\ref{sec:spmod})
because of the low counting statistics of the spectra). We also
calculated the hardness ratio, defined as the count rate in 1--8 keV
divided by that in 0.3--1 keV and obtained the values of 0.71$\pm$0.11
and 0.76$\pm$0.23 for the bright and faint intervals,
respectively. Thus we see no clear spectral variability associated
with the flares in observation C1.

\subsection{Spectral Modeling}
\label{sec:spmod}
\tabletypesize{\tiny}
\setlength{\tabcolsep}{0.02in}
\begin{deluxetable*}{r|c|c|l|c|cc}
\tablecaption{Spectral fit results \label{tbl:spfit}}
\tablewidth{0pt}
\tablehead{Obs ID & model & $N_{\rm H}$ &  Other Parameters & $\chi^2_\nu(\nu)$\tablenotemark{a} & $L_{\rm abs}$ & $L_{\rm unabs}$\\
             &  & (10$^{22}$ cm$^{-2}$)&         &    & \multicolumn{2}{c}{(10$^{38}$ erg s$^{-1}$)}\\
(1)&(2)&(3)&(4)&(5)&(6)
}
\startdata
0094360601(X1) & PL & 0.06 & $\Gamma_{\rm PL}$=2.5 & \nodata & 0.22$\pm$0.06 & 0.4$\pm$0.1\\
\cline{1-7}
0094360701(X2) & PL & 0.06 & $\Gamma_{\rm PL}$=2.5 & \nodata & 0.37$\pm$0.09 & 0.6$\pm$0.2\\
\cline{1-7}
\multirow{5}{*}{0404980101(X3)} & PL & $0.15^{+0.01}_{-0.01}$ & $\Gamma_{\rm PL}$=$2.37^{+  0.04}_{ -0.04}$, $N_{\rm PL}$=$22^{+  1}_{ -1}\times10^{-5}$ &$1.31(560)$ & $21.5^{+ 0.5}_{-0.5}$  & $42.2^{+ 2.0}_{-1.9}$ \\
\cline{2-7}
& MCD & $0.01^{+0.01}$  & $kT_{\rm MCD}$=$ 0.86^{+ 0.02}_{-0.02}$, $N_{\rm MCD}$=$5.5^{+0.5}_{-0.4}\times10^{-2}$ &$1.15(560)$ & $19.9^{+ 0.5}_{-0.5}$  & $20.6^{+ 0.5}_{-0.5}$ \\
\cline{2-7}
&MCD+PL & $0.04^{+0.02}_{-0.02}$  & $kT_{\rm MCD}$=$ 0.87^{+ 0.06}_{-0.06}$, $N_{\rm MCD}$=$3.6^{+1.2}_{-0.9}\times10^{-2}$, $\Gamma_{\rm PL}$=$2.1^{+  0.2}_{ -0.2}$, $N_{\rm PL}$=$6^{+2}_{-2}\times10^{-5}$ &$0.95(558)$ & $21.2^{+ 0.6}_{-0.5}$  & $25.5^{+ 2.3}_{-1.8}$ \\
\cline{2-7}
& MCD+nthComp & $0.02^{+0.01}$ &  \parbox[t]{3in}{$kT_{\rm MCD}$=$0.45^{+0.15}_{-0.15}$, $N_{\rm MCD}$=$0.23^{+0.16}_{-0.23}$, $\Gamma_{\rm nthComp}$=$1.7^{+0.5}_{-0.7}$, $kT_{\rm e,nthComp}$=$1.1^{+0.3}_{-0.2}$, $N_{\rm nthComp}$=$9^{+7}_{-8}\times10^{-5}$} &$0.90(557)$ & $20.6^{+ 0.5}_{-0.5}$ & $21.5^{+ 0.7}_{-0.7}$ \\
\cline{2-7}
& MCD+CompTT & $0.01^{+0.01}$ & \parbox[t]{3in}{$kT_{\rm MCD}$=$0.25^{+0.07}_{-0.04}$, $N_{\rm MCD}$=$1.7^{+0.7}_{-0.7}$, $\tau_{\rm compTT}$=$11^{+1}_{-1}$, $kT_{\rm e,compTT}$=$1.0^{+0.1}_{-0.1}$, $N_{\rm compTT}$=$2.8^{+0.6}_{-0.7}\times10^{-4}$} & $0.90(557)$ & $20.6^{+ 0.5}_{-0.5}$ & $21.1^{+ 0.7}_{-0.5}$\\
\cline{1-7}
\multirow{2}{*}{808(C1)} &PL & $0.06^{+0.07}$ & $\Gamma_{\rm PL}$=$  2.5^{+  0.5}_{ -0.4}$, $N_{\rm PL}$=$8^{+  2}_{ -2}\times10^{-6}$ &$0.77( 10)$ & $ 0.9^{+ 0.2}_{-0.2}$  & $ 1.5^{+ 0.8}_{-0.3}$ \\
\cline{2-7}
 &MCD & $0.01^{+0.02}$ & $kT_{\rm MCD}$=$ 0.43^{+ 0.08}_{-0.06}$, $N_{\rm MCD}$=$3.3^{+3.1}_{-1.7}\times10^{-2}$ &$1.92( 10)$ & $ 0.7^{+ 0.1}_{-0.1}$  & $ 0.7^{+ 0.1}_{-0.1}$ \\
\cline{1-7}
\multirow{2}{*}{9553(C2)} &MCD & $0.01^{+0.05}$  & $kT_{\rm MCD}$=$ 0.68^{+ 0.10}_{-0.09}$, $N_{\rm MCD}$=$2.6^{+2.1}_{-1.0}\times10^{-2}$ &$0.71( 11)$ & $ 3.6^{+ 0.3}_{-0.5}$  & $ 3.7^{+ 0.4}_{-0.4}$ \\
\cline{2-7}
 & PL & $0.24^{+0.14}_{-0.11}$ & $\Gamma_{\rm PL}$=$2.9^{+0.5}_{ -0.4}$, $N_{\rm PL}$=$6^{+3}_{-2}\times10^{-5}$ &$0.42( 11)$ & $ 3.7^{+ 0.6}_{-0.5}$  & $14.4^{+ 14.0}_{-5.6}$ \\
\cline{1-7}
\multirow{2}{*}{RP600050N00(R1)} &MCD & $0.01^{+0.01}$  & $kT_{\rm MCD}$=$0.42^{+ 0.04}_{-0.04}$, $N_{\rm MCD}$=$0.46^{+0.21}_{-0.13}$ &$1.07( 16)$ & $ 8.3^{+ 0.5}_{-0.5}$  & $ 8.9^{+ 0.5}_{-0.5}$ \\
\cline{2-7}
& PL & $0.03^{+0.01}$ & $\Gamma_{\rm PL}$=$2.1^{+  0.2}_{ -0.2}$,  $N_{\rm PL}$=$9.2^{+  0.6}_{ -0.6}\times10^{-5}$ &$1.08( 16)$ & $15.0^{+ 2.6}_{-1.9}$  & $18.3^{+1.6}_{-1.2}$ \\
\cline{1-7}
RH600678N00(R2) & PL & 0.06 & $\Gamma_{\rm PL}$=2.5 & \nodata & 0.2$\pm$0.2 & 0.4$\pm$0.4 \\
\cline{1-7}
RH600769N00(R3) & PL & 0.06 & $\Gamma_{\rm PL}$=2.5 & \nodata & 0.0$\pm$0.5 & 0.0$\pm$0.8
\enddata 
\tablecomments{The fits were carried out only on observations X3, C1, C2, and R1. The 0.2-10 keV absorbed ($L_{\rm abs}$) and unabsorbed ($L_{\rm unabs}$) luminsosities are given. For observations X1, X2, R2, and R3, they were estimated based on the PL fit to observation C1. The energy bands of the fits are 0.2--10 keV for X3, 0.3--8 keV for C1 and C2, and 0.2--2.4 keV for R1. All errors are at a 90\%-confidence level.} 
\end{deluxetable*}

The brightest observation X3 has the best quality and allows for
relatively detailed modeling. Following \citet{strowi2006}, we fitted
the 2--10 keV spectrum with a PL and a broken PL (both unabsorbed) and
found that the latter fit (the total $\chi^2$ is 98.2 for 118 d.o.f.)
showed a significant ($5\sigma$) improvement over the former one (the
total $\chi^2$ is 125.2 for 120 d.o.f.), indicating the presence of
spectral curvature at several keV (the inferred break energy is
$4.6\pm0.7$ keV). The fits over the whole spectrum (0.2--10 keV) with
a PL is very poor, with a reduced $\chi^2$ of 1.31 (560 d.o.f.,
Table~\ref{tbl:spfit}). The fit with the MCD model is better (the
reduced $\chi^2$ is 1.15), but there are still clear residuals at high
energies, as is often seen in the spectral fits to ULXs and was
explained as due to the advection effect and/or the Compton scattering
of disk photons in an optically thick wind, photosphere or corona in
many studies, such as \citet{glrodo2009} and \citet{misuro2012,
  mimima2013}. Similar to these studies, we tested two-component
models to include a Comptonization component, i.e. MCD+PL and
MCD+CompTT (CompTT \citep{ti1994} is available in XSPEC). The fit with
the MCD+PL model is much better than that with the MCD model,
decreasing the total $\chi^2$ by 114. The MCD+CompTT model, with the
seed photons tied to the inner disk temperature, decreases the total
$\chi^2$ further, by 32 compared with the MCD+PL model for one more
d.o.f. The MCD+CompTT model inferred a population of cool (1.0 keV)
Comptonizing electrons with a high optical depth ($\tau\sim11$). For
the disk, we inferred $kT_{\rm MCD}=0.25_{-0.04}^{+0.07}$ keV (90\%
error) or $kT_{\rm MCD}=0.25_{-0.05}^{+0.30}$ keV ($2\sigma$ error)
(here we report two error bars because we found that $\chi^2$ varied
slowly for a relatively large range of $kT_{\rm MCD}$).

We also tested the Comptonization model nthComp \citep[in
  XSPEC,][]{zydosm1999,zdjoma1996,lizd1987}, whose seed photons can be
assumed to have the MCD shape. This model was used in the fits to very
bright (around the Eddington limit) spectra from the $\rho$ state of
the BHB \object{GRS 1915+105} by \citet{nerele2011} and those from the
luminous neutron-star (NS) low-mass X-ray binary (LMXB) \object{GX
  17+2} by \citet{lireho2012} as a substitution of the SIMPL model
\citep{stnamc2009} to account for the high-energy cutoff in the
spectra. The fitting results with MCD+nthComp model are given in
Table~\ref{tbl:spfit} and the unfolded spectrum is shown in
Figure~\ref{fig:indlcsp}. The MCD normalization is not well
constrained, with the 90\%-confidence lower limit reaching zero, which
was also seen in \citet{nerele2011} and \citet{lireho2012}. Following
\citet{lireho2012}, we roughly estimated the pre-Comptonization disk
normalization $N_{\rm disk, pre-Compton}$=$(R_{\rm in, km}/D_{\rm
  10kpc})^2\cos i$, where $R_{\rm in, km}$ is the inner disk radius,
$D_{\rm 10kpc}$ the source distance, and $i$ the disk inclination, by
adding the photons Compton scattered (the nthComp component) and those
unscattered (the MCD component), assuming that the photons are
conserved. We obtained $N_{\rm disk,
  pre-Compton}$=$0.47_{-0.23}^{+0.91}$. The optical depth of the
corona, which is not an explicit parameter of the model but can be
derived using Equation A1 in \citet{zdjoma1996}, is $\sim$22, which is
high, similar to that inferred from the MCD+CompTT model. Therefore we
have made the simple assumption that while the optically thick corona
can have seed photons from the disk it does not strongly affect the
thermal disk emission in turn in both the MCD+CompTT and MCD+nthComp
models.

The other spectra have much lower quality, and we only test the
single-component models, i.e., a MCD and a PL. The unfolded spectra
are plotted in Figure~\ref{fig:indlcsp}. The {\it ROSAT}/PSPCB
observation R1 can be fitted with either model. We caution that the
fit for this observation is limited by the narrow energy band coverage
(0.2--2.4 keV) and calibration uncertainty. The {\it Chandra}
observation C2 can also be fitted with either model, but the PL model
requires a relatively strong absorption ($N_{\rm
  H}$=0.24$\times10^{22}$ cm$^{-2}$) and a steep PL ($\Gamma_{\rm
  PL}$=2.9). Observation C1, the one with (quasi-)periodic flares
(Figure~\ref{fig:indlcsp}), can be fitted with a PL, but not well with
the MCD model (the reduced $\chi^2$ is 1.9). The inferred photon index
is $\Gamma_{\rm PL}=2.5^{+0.5}_{-0.4}$.

We did not carry out spectral fits to the other four observations (R2,
R3, X1, and X2) when the source was not significantly detected and
subject to strong contamination from bright sources near the
nucleus. Instead, we estimated their source luminosities assuming the
best-fitting PL model from observation C1, since C1 was the closest in
flux to these observations (Table~\ref{tbl:spfit}). We note that for
observations R2 and R3, in which the source was not detected, we
calculated the 90\% confidence intervals using Bayesian statistics
\citep{krbuno1991}.

\subsection{The Optical Counterpart} 

The candidate optical counterpart to ULX2 fortunately appears on the
outskirts of a star-forming region devoid of bright
sources. \verb|HSTPhot| indicates this counterpart as point-like in
all filters. Its VEGA magnitudes are given in Table~\ref{tbl:hst}. The
counterpart seems relatively red. It was the most significantly
detected in the F555W and F814W filters. Further considering that
these two filters are less subject to possible emission from accretion
activity than the other filters, we used the SYNPHOT package to
compare the color of these two filters with the stellar spectra in
\citet{pi1998}, assuming the Galactic extinction of $E$(B$-$V)=0.02
\citep{scfida1998}. We found that the counterpart has the
F555W$-$F814W color (1.21 mag) the closest to those of G8I (1.09 mag)
and K3V (1.16 mag) stars among supergiants and dwarfs in
\citet{pi1998}, respectively. We fitted the F450W, F555W, and F814W
fluxes of the counterpart with these two stellar spectra by minimizing
the total $\chi^2$, with the normalization of the spectra as a free
parameter. The corresponding apparent magnitudes of the best-fitting
spectra are listed in Table~\ref{tbl:hst}. The fit residuals are
$\lesssim$0.1 mag in these filters. In the $H_\alpha$ filter F656N,
the deviation is slightly larger, with our source brighter than the
fits by 0.3--0.4 mag (2--3$\sigma$). In the UV filter F336W, the
excess is much larger, by 1.5 and 0.7 mag compared with the fits with
a G8I star and a K3V star, respectively, though in this filter the
source was detected only at 6$\sigma$. The apparent visual magnitude
of our source is $V$=22.22, corresponding to an absolute magnitude of
$-6.4$ at the distance of M94, which is close to that expected for a
G8I star or that for a red globular cluster containing
$\sim$$2\times10^5$ K3V stars. In the above, we have neglected the
possible effect of binary evolution on the color of the donor star.

\section{DISCUSSION}
\label{sec:dis}
\subsection{ULX2 as a Super-Eddington Accreting Stellar-mass BHB}
ULX2 was fortunately captured at different flux levels, showing clear
spectral evolution. With similar spectral models tested, we can easily
compare ULX2 with the second ULX in M31 (\object{XMMU
  J004243.6+412519}) studied by \citet{mimima2013}. The brightest
observation X3 of ULX2 is very similar to the brightest {\it
  XMM-Newton} observations XMM3 and XMM4 of \object{XMMU
  J004243.6+412519}. All these observations have $L_{\rm X}\sim 10^{39}$
erg~s$^{-1}$ and are better fitted by the MCD+CompTT model than by a
PL, a MCD, or their combination. The fits with the MCD+CompTT model to
these observations all inferred a relatively cool ($<$1 keV, which was
not well constrained) disk and Comptonization in a cool ($\sim$1 keV)
optically thick ($\tau\sim10$) corona. \object{XMMU J004243.6+412519}
is probably a stellar-mass BHB with $M_{\rm BH}\sim10$ \msun, based on
the joint radio/X-ray behavior and the observation of a disk-dominated
state at a very low luminosity, supporting that the above spectra are
probably characteristic of accretion at the Eddington limit
\citep{mimima2013}. Then ULX2 could be a stellar-mass BHB as well,
with an accretion rate around the Eddington limit in observation
X3. Many low-luminosity ULXs show similar spectra and could also be
explained as supercritically accreting stellar-mass BHBs
\citep{mimima2013}.

If ULX2 is really a stellar-mass BHB, we would expect it to behave
similar to typical Galactic BHBs when it is at well sub-Eddington
luminosities. In the fainter observation C2 ($L_{\rm
  X}\sim3.6\times10^{38}$ erg~s$^{-1}$), the spectrum is softer than
observation X3 and can be described with a standard thermal accretion
disk, thus consistent with being in the thermal state of Galactic
BHBs. The thermal state in Galactic BHBs tends to occur above
$\sim3\%$ of the Eddington limit \citep{dufeko2010}. Assuming
observation C2 to be in this state and using its 0.2--10 keV
unabsorbed luminosity from the MCD model, we constrained the mass of
the BH in ULX2 to be $\lesssim$100 \msun, supporting the
identification as a stellar-mass BHB. We note that for this
observation we cannot rule out a steep PL model with relatively strong
absorption from the fit. In the even fainter observation C1 ($L_{\rm
  X}\sim9\times10^{37}$ erg~s$^{-1}$), the spectrum becomes harder
again and can be described with a PL with $\Gamma_{\rm
  PL}\sim2.5\pm0.5$. Such a photon index is often seen in the steep-PL
state of Galactic BHBs, and the source might be in such a state in
this observation, but considering its large uncertainty, we cannot
rule out the source being in the hard state instead. In any case, we
might have observed a state transition often seen in BHBs. We clearly
need higher quality data for more detailed comparison.

Therefore transient/highly variable ULXs like ULX2 and \object{XMMU
  J004243.6+412519} serve a link between persistent low-luminosity
ULXs and classical stellar-mass BHBs. This is reminiscent of the first
known transient Z source \object{XTE J1701-462} linking Z and atoll
sources, the two main classes of weakly magnetized NS LMXBs
\citep{lireho2009,hovafr2010}. Here we briefly discuss the NS LMXB
case to gain some insights into the physics involved in the BHB
case. Z sources reach the Eddington limit or above and are mostly
persistent with variability factors of only a few, while atolls have
luminosities typically less than 50\% of the Eddington limit and tend
to be transient or highly variable. The proof that Z sources accrete
at near or super-Eddington limit and is the same class of object as
atolls but at different accretion rates based on \object{XTE
  J1701-462} is very convincing in several ways
\citep{lireho2009,hovafr2010}. First, Z and atoll sources have
well-known distinct timing and spectral properties, and \object{XTE
  J1701-462} showed Z-source properties at high luminosities and then
atoll-source ones during the decay in its 2006-2007
outburst. Secondly, NS LMXBs have a unique way to infer the Eddington
luminosity, i.e. radius expansion Type-1 X-ray bursts, which were
detected in \object{XTE J1701-462}, and the luminosities when
\object{XTE J1701-462} behaved as a Z source were indeed near or above
the Eddington luminosity inferred from the bursts \citep{lialho2009}.
Finally, the spectral fitting by \citet{lireho2009} also gathered some
evidence that the Eddington limit was reached in the Z-source stage,
especially the result that the disk in the Z-source lower vertex had a
relatively constant temperature and an increasing inner disk radius
with increasing luminosity, deviating from the $L \propto T^4$ trend
observed in the atoll-source stage. The increase in the inner disk
radius with luminosity is expected because as the mass accretion rate
into the disk increases, more and more inner part of the disk reaches
the local Eddington limit, leading to increasing radial advection flow
and/or mass outflow \citep{ohmi2007}.

Therefore studies of NS LMXBs not only demonstrate that near or
super-Eddington accretion is possible but also assure us that objects
with different outbursting behaviors and luminosities can belong to
the same class and have overall spectral/timing properties mainly
determined by the accretion rate. All this supports the possibility
that most low-luminosity ULXs are in fact near or super-Eddington
accreting stellar-mass BHBs. Galactic BHBs are known to show many
similarities to atolls, such as sub-Eddington luminosity, transient
behavior, broad-band noise, and possibly also disk spectral evolution
\citep{va2006,lireho2007}, though there are also some differences,
such as hotter thermal spectra and stronger millisecond variability in
atolls, which could be reasonably ascribed to emission from the impact
of materials onto the NS surface. Then there is a question of whether
ULXs that are supercritically accreting stellar-mass BHBs also show
some properties observed in Z sources. One possible interesting
similarity is their generally small long-term variability. Classical Z
sources (\object{Sco X-1}, \object{GX 17+2}, \object{GX 349+2},
\object{GX 340+0}, \object{GX 5-1}, and \object{Cyg X-2}), all with a
low-mass companion, are persistent with long-term variation factors of
only a few (only two Z sources (\object{XTE J1701-462} and \object{IGR
  J17480-2446}), discovered recently, are transients). Most known ULXs
are also persistent. \citet{liweba2012} found only 15 with a long-term
variation factor $>$10 in their 100 ULXs, which is in contrast with
the general transient behavior of Galactic BHBs, especially those with
a low-mass companion. This could be because most of these ULXs might
have a high-mass companion, but from Z sources we cannot rule out that
the long-time activity might be related to high accretion rates. It
would also be worthwhile to search for similarity of the disk behavior
between ULXs and Z sources. \citet{nerele2011} had reported similar
disk evolution in the BHB \object{GRS 1915+105} in a very bright
state, i.e., the $\rho$ state, to that seen in \object{XTE J1701-462}
in the Z-source stage. A lot of work is still needed in the future to
establish the possible connection between Z sources and low-luminosity
ULXs.

\subsection{Cause of Short-term X-ray Variability}
The observations of dips from ULX2 make it one of the dipping sources,
which include about 20 Galactic X-ray binaries and a few ULXs, such as
the ULX in NGC 55 \citep{strowa2004} and NGC 5408 X-1
\citep{past2013,grkaco2013}. Long dips lasting for 10\%--30\% of the
orbital phase are mostly found in NS LMXBs and can be explained as due
to absorption by a bulge on the edge of the accretion disk at the
point where the gas stream impacts the disk
\citep[e.g.,][]{whsw1982}. Short dips less than a few hundred seconds
(thus typically $<$1\% of the orbital period) are commonly observed in
two BHBs \object{GRO J1655-40} (with a low-mass companion) and
\object{Cyg X-1} (with a supergiant companion). They show orbital
phase dependence and could be due to absorption in accretion streams
from the companion \citep{kuinco2000,bachch2000,fecu2002}. It is not
clear whether the dipping ULXs in NGC 55 and NGC 5408 also show
orbital phase dependence, but they might have similar origin. Dips in
ULX2 are short and are thus probably also due to absorption in
accretion streams. The spectrum becomes soft in the dips in
observation X3, which could be because a hot component is obscured
and/or because the absorbing matter is partially ionized. In terms of
the MCD+nthComp model that we used to fit observation X3, the dipping
spectrum that we created from this observation could be very roughly
accounted for with either complete obscuration of the nthComp
component (the reduced $\chi^2$ is 3.7 for 7 d.o.f.) or with the
presence of absorbing matter with a column density of
$6.5\times10^{23}$ cm$^{-2}$ and ionization parameter of $\log\xi=2.9$
(the reduced $\chi^2$ is 4.3 for 7 d.o.f., using the ionized
absorption model zxipcf in XSPEC). We do not have enough dips to
search for periodicity, which could otherwise provide information of
the orbital period. In any case, the dips often imply that the system
is at a high inclination. Detection of ULXs at high inclinations is
important, because it poses a problem to use the beaming effect to
explain the ultraluminous nature of these sources.

The possible large (quasi-)periodic X-ray modulations/flares in
observation C1 is also a special property of ULX2. Similar modulations
were also observed in some ULXs such as \object{CXOU J141312.3-652013}
in Circinus \citep[$\sim$7.5 hr,][]{babrsa2001}, \object{M51 X-7}
\citep[$\sim$2 hr,][]{librir2002}, and \object{CXOU J013651.1+154547}
in M74 \citep[$\sim$2 hr,][]{krkiga2005, librll2005}. The main
difference is that ULX2 has $L_{\rm X}\sim 10^{38}$ erg~s$^{-1}$ while
those sources have $L_{\rm X} > 10^{39}$ erg s$^{-1}$ when such large
modulations were observed. Various explanations for such modulations
were discussed by those studies, including the binary eclipse,
modulations of the accretion rate due to some instability, and
variability at the base of the jet. Because ULX2 might have a high
inclination considering the detection of dips, if the flares are
periodic, the binary eclipsing explanation seems plausible, with the
$\sim$7-hr period being the orbital period. The main problem for this
explanation is that we found no spectral change accompanying the flux
variation. Moreover, such modulations seem to be flux and/or spectral
state dependent because they were not seen in the brighter/softer
observations. Therefore the disk or jet instability explanations are
favored.

\subsection{Nature of the Optical Counterpart}
Another interesting aspect of ULX2 is our identification of its
point-like red optical counterpart candidate. Some ULXs have been
reported to have optical counterpart candidates
\citep[e.g.,]{tafegr2011,glcohe2013}. Most of them appear blue and are
probably contaminated by accretion activity, while some have a clear
red component like our source, such as \object{IC 342 X-1}
\citep{feka2008} and \object{M81-ULS1} \citep{lidi2008}. For
\object{M81-ULS1}, there is also an extra blue component which was
shown by \citet{lidi2008} to be probably from disk emission, while the
red component could be an asymptotic giant branch star. There is some
possible UV excess in ULX2, and the most likely explanation is the
disk emission too. We note that our source is highly variable while
the {\it HST} observations are not simultaneous with each other or
with X-ray observations. Therefore, one explanation for why the
counterpart to ULX2 appears red while those to other ULXs are mostly
blue is that the {\it HST} observations of ULX2 were made when it was
not X-ray ultraluminous, though the alternative explanation that ULX2
resides in an old cluster is also possible.

\section{CONCLUSIONS}
\label{sec:conclusion}
We have shown many intriguing properties of ULX2. It is either a
highly variable persistent source or a recurrent transient, with a
long-term variation factor of at least $\sim$100 and a duty cycle of
at least $\sim$70\%. In the brightest observation X3, the source
exhibited as a member of low-luminosity ULXs, with $L_{\rm
  X}\sim2\times10^{39}$ erg~s$^{-1}$ and had a spectrum showing a
high-energy cutoff but too hard to be described by thermal emission
from a standard accretion disk. Short dips, accompanied by spectral
softening, are also clearly seen. In the fainter observation C2
($L_{\rm X}\sim3.6\times10^{38}$ erg~s$^{-1}$), the spectrum became
softer and is consistent with thermal emission from a standard
accretion disk with $kT_\mathrm{MCD}\sim0.7$ keV, supporting it as in
the thermal state as seen in Galactic BHBs. In the even fainter
observation C1 ($L_{\rm X}\sim9\times10^{37}$ erg~s$^{-1}$), the
spectrum becomes harder again, with $\Gamma_{\rm PL}\sim2.5\pm0.5$,
and could be in the steep-PL state or the hard state of Galactic
BHBs. In this observation, we also observe possible flares that might
indicate $\sim$7-hr periodic X-ray modulations. We also identify a
possible point-like optical counterpart, whose colors and luminosity
resemble a G8 supergiant or a compact red globular cluster containing
$\sim2\times10^5$ K dwarfs except for some possible UV
excess. Combining all the above properties, we suggest that ULX2 is a
stellar-mass BHB with a supergiant companion or with a dwarf companion
residing in a globular cluster. The high variability and large duty
cycle of ULX2 makes it easy for multiwavelength follow-up, which is
important for confirmation of some properties that we report here,
such as the state transition and possible periodic behavior at low
luminosities, and reveal more properties to further constrain its
nature.

\acknowledgments Acknowledgments: We thank the anonymous referee for
the helpful comments. This work is supported by NASA Grant NNX10AE15G.

\end{document}